\documentclass[useAMS,usenatbib]{mn2e}
\usepackage{graphicx}
\usepackage[all]{xy}
\usepackage{amsmath}
\usepackage{amssymb}

\title[Probing SSs with Adv. Grav. Wave Detectors]
  {Probing Strange Stars with Advanced Gravitational Wave Detectors}
\author[P.H.R.S. Moraes and O.D. Miranda]
  {Pedro H.R.S. Moraes,$^1$ Oswaldo D. Miranda$^1$
  \newauthor 
  $^1$INPE - Instituto Nacional de Pesquisas Espaciais - Divis\~ao de Astrof\'isica \\
Av. dos Astronautas 1758, S\~ao Jos\'e dos Campos, 12227-010 SP, Brazil}
\date{Released 2014 Xxxxx XX}

\pagerange{\pageref{firstpage}--\pageref{lastpage}} \pubyear{2014}

\def\LaTeX{L\kern-.36em\raise.3ex\hbox{a}\kern-.15em
    T\kern-.1667em\lower.7ex\hbox{E}\kern-.125emX}

\begin{document}

\label{firstpage}

\maketitle

\begin{abstract}
 When a neutron star is compressed to huge densities, it may be converted to a strange star. In property of the event/year rate of a neutron star - strange star binary system, we show that the operational phase of advanced gravitational wave detectors may bring up some evidences that such strange stars do exist. Moreover we argue that such a system could be a plausible progenitor to GRB 051103 and GRB 070201, whose non-detection by LIGO last run awaits convincing explanation.
 \end{abstract}

\begin{keywords}
 gravitational waves --  binaries -- gamma-ray bursts -- neutron star.
\end{keywords}

\section{Introduction}
A neutron star (NS) is a supernova explosion remnant known for its high density, strong gravitational field and rapid rotation rate. Neutron star binary systems (NSBSs) are among the leading potential sources for gravitational wave (GW) detection (\cite{phinney/1991, cutler/1993}). They also have an expressive observational counterpart since they are related to the most extreme explosive events in the Universe, the gamma ray bursts (GRBs). Currently, they are the leading candidate to explain short-duration gamma ray bursts (sGRBs) - those whose bursts last less than $2$s (\cite{kouveliotou/1993}) - which could also be generated by a NS - black hole (BH) binary system (BS) (\cite{nakar/2007}). 

The most accurate measurements concerning NSs are mass determinations from pulsar timing. Although most NSs have masses close to $1.3-1.4\,{\rm M}_\odot$ (\cite{lattimer/2012}), there is ample observational support from pulsars for NSs with masses significantly greater than $1.4\,{\rm M}_\odot$, e.g., the PSR J$1614-2230$ with mass $1.97\pm0.04\,{\rm M}_\odot$ (\cite{guillemot/2012}), which causes some controversy about their origin. Depending on the star's mass and rotational frequency, the matter in the core regions of NSs may be compressed to huge densities, up to an order of magnitude greater than that on atomic nuclei (\cite{weber/2007}). When nucleon matter is squeezed to a sufficiently high density, it turns into a uniform two-flavor - 'up and down' - matter. However, this matter is unstable and consequently it is converted to a three-flavor - 'up, down and strange' - quark matter, named strange matter (SM). Although one still does not know precisely from experimentation at what density the phase transition to SM occurs (\cite{weber/2006}), it is believed that neutrons inside such NSs can undergo a transition to their constituent quarks, resulting in a quark star. Weak interactions convert about a third of the up and down quarks to strange quarks (\cite{koshy/2012}), thus allowing these stars to be also referred as strange stars (SSs).

The SM hypothesis implies that an SS has a very different {\it mass vs. radius} relation when compared to an NS, since they have different equations of state (EoS). From the conventional Akmal-Pandharipande-Ravenhall EoS for NSs (\cite{akmal/1998}) one can note that it implies a lower mass limit of $0.08 \,{\rm M}_\odot$. On the other side, SSs have no theoretical lower mass limit. The inexistence of a lower mass limit for SSs comes from the strong possibility that SM is more stable that ordinary nuclear matter (\cite{bodmer/1971, witten/1984}). This, in the physical conditions of a compact object interior, may be what prevents the star from collapsing, producing a more compact object. It is very essential to try to identify this object and distinguish it from an NS. An NS with mass $\sim 0.2\,{\rm M}_\odot$, for instance, has a radius $R>15\,{\rm km}$, whereas for an SS with the same mass, $R\lesssim 5\,{\rm km}$ (\cite{xu/2005}). The distinction between NSs and SSs, in this way, can be done by measuring the radii of low-mass pulsar-like stars by X-ray satellites.

Another way of distinguishing NSs from SSs could rise from the GW detection (\cite{baus/2010}). With the current and upcoming GW detectors like Advanced LIGO (aLIGO) (\cite{harry/2010}) and Einstein Telescope (ET) (\cite{sathyaprakash/2012}), one could theorize if the signal from coalescing NSs and SSs could be distinct one from another. Compact object binaries are among the most promising sources of GWs for these detectors (\cite{nakar/2007, blinnikov/1984}). Also, from GWs emitted during the moments that precede the last stable orbit of the BS (i.e., right before coalescence) could raise essential information about the EoS of dense matter (\cite{faber/2002, taniguchi/2003, oechslin/2004, bejger/2005}).

However, when studying GWs emitted by a compact BS, an important unresolved issue arises. \cite{abbott/2008} and \cite{abadie/2012} have analyzed GRB 070201 and GRB 051103, respectively, from LIGO data and found no evidence of GW signal. Since the above GRBs are believed to be inside the horizon distance that LIGO could observe for this kind of systems, it implies that NS and NS-BH BSs can be excluded as plausible progenitors. There are still some uncertain and/or unsolved questions concerning these systems apart from what is mentioned above. \cite{berger/2010}, for instance, presented the ``no-host'' problem for several sGRBs which present no bright coincident host galaxies. \cite{coward/2007} presented some open issues with the GRB redshift distribution, focusing on problems of constraining GRB rate evolution. Also, the magnetic field generation in GRBs is still unsolved (\cite{fiore/2006}). The origin of the required magnetic field is under discussion (\cite{coburn/2003, waxman/2003}) and the Weibel instability (\cite{weibel/1959}) seems to be the leading candidate to explain it (\cite{medvedev/1999, medvedev/2005}). For a broad review on GRB problems, see (\cite{zhang/2004}).

It is important to mention that the SM could also be processed inside white dwarfs (WDs) by the accretion of a SM nugget (\cite{glendenning/1995a, glendenning/1995b}) ejected from the coalescence of SM NSs, as it is shown in \cite{mathews/2006}. Once captured by a WD, SM nuggets gravitationally settle to its center and begin to convert normal matter into SM, what eventually lead to the formation of an extended SM core in the WD. Once SM can be more compact than electron degenerate matter, strange dwarfs must have smaller radius than WDs.

In particular, \cite{mathews/2006} have compared the masses and radii of 22 nearby WDs collected from \cite{provencal/1998, provencal/2002} with the standard mass-radius relation (\cite{hamada/1961}), which has led to some evidence for the existence of two WDs populations, one significantly more compact than the other. One could suggest that the issues pointed out above could be renewed with the possible existence of SSs. Our aim in this work is to investigate the GW sign of an NS-SS BS and we have clarified the importance of considering the existence of SSs, as well as discussed the possibility of such systems being the progenitors of GRB 070201 and GRB 051103.

\section{An NS-SS coalescing binary as source of gravitational waves}

As it was already cited, most NS masses are in the range $1.3-1.4$M$_\odot$. In this way, let us investigate the GW sign of an NS-SS BS with masses $m_{NS}=m_1=1.3$M$_\odot$ and $m_{SS}=m_2=0.7$M$_\odot$. When analysing the radii of these stars for different EoS such as the one presented in \cite{haensel/1986}, one can notice that $R^{SS}<R^{NS}$ in such a way we can use, for the last stable orbit of this coalescing system, the same equation which is used for an NSBS (\cite{sathyaprakash/2001}), i.e., $f_{LSO}=1.5(\tilde{M}/2.8M_\odot)^{-1}\,{\rm kHz}$, with $\tilde{M}=M(1+z)$ being the total redshifted mass of the NSBS. Note that the same approach has been fulfilled by \cite{regimbau/2006}, showing that an NSBS could reach frequencies up to $\sim 1.5\,{\rm kHz}$. One can evaluate the luminosity distance that this system can reach from \cite{arun/2005}:

\begin{equation}\label{eqn:2}
d_L(M,\eta)=\frac{1}{\rho _0\pi^{2/3}}\left[\frac{2{\eta\tilde{M}^{5/3}}}{15}\int_{f_s}^{f_{LSO}}\frac{f^{-7/3}}{S_h(f)}df\right]^{1/2}.
\end{equation}
In Eq.(\ref{eqn:2}), $\rho_0$ is the signal-to-noise ratio achieved in the GW detector, ${\eta}=m_1m_2/{M}^{2}$ is the dimensionless mass ratio, $S_h(f)$ is the analytical fit of the noise curve for the GW detector, $f_s$ is a low-frequency cutoff below which $S_h(f)$ can be considered infinite and the integration is computed through the frequency $f$. From \cite{mishra/2010}, we obtain the noise spectral density $S_h(f)$ for aLIGO and ET in order to evaluate Eq.(\ref{eqn:2}), which assuming $\rho_0=8$, results in $d_L^{aLIGO}\sim 145\,{\rm Mpc}$ and $d_L^{ET}\sim 2.5\,{\rm Gpc}$.

With the knowledge of the NS-SS BS luminosity distance for both detectors, one can estimate their event/year rate $N$. In \cite{corvino/2012}, a method is presented to obtain $N$ from the Galactic merger rate $\mathcal{R}$ of the system and the detectable volume $V_d$ of the concerned source. From \cite{kalogera/2004,belczynski/2008,abadie/2010}, $\mathcal{R}^{NS-NS}\sim 10^{-4}{\rm yr}^{-1}$ while from \cite{bauswein/2009}, $\mathcal{R}^{SS-SS}\sim 10^{-5}-10^{-4}{\rm yr}^{-1}$. In this way, since $\mathcal{R}^{NS-NS}\sim 10\mathcal{R}^{NS-BH}$ which in turn is $\sim10\mathcal{R}^{BH-BH}$ (see Table II of \cite{abadie/2010}), it is conservatively plausible to assume that $\mathcal{R}^{NS-SS}$ is somewhere between $\mathcal{R}^{NS-NS}$ and $\mathcal{R}^{SS-SS}$, which agrees with the fact that these events are sufficiently rare so that they were still not detected. Note that, \cite{belczynski/2002} have shown, 
through the use of population synthesis method, that if quark stars exist at all, their population can be as large as the population of BHs. Thus, we can also think that $\mathcal{R}^{NS-BH}\sim \mathcal{R}^{NS-SS} \lesssim 10^{-5}{\rm yr}^{-1}$. Therefore taking $\mathcal{R}^{NS-SS}=10^{-5}{\rm yr}^{-1}$ we can calculate the event/year rate for these systems from \cite{corvino/2012}:

\begin{equation}\label{eqn:3}
N=V_d\times 0.0116\times \frac{1}{2.26^{3}}\times\mathcal{R},
\end{equation}
with $V_d=(4/3)\pi d^{3}_L$ being the detectable volume. The numerical factor $0.0116$ is an estimate of the local number density of Milky Way equivalent galaxies in ${\rm Mpc}^{-3}$ (\cite{kopparapu/2008}) and the factor $1/2.26^{3}$ is needed to average over sky position and orientation. From Eq.(\ref{eqn:3}) and the luminosity distance values obtained for both detectors, we find $N^{aLIGO}\sim 0.13$ events per year and $N^{ET}\sim 700$ events per year.

\section{The uncertainty on the binary system masses}

In order to guarantee that we have an SS in the concerned BS, it is fundamental to calculate the uncertainty on the masses of the stars. A reliable method to do so is through the Fisher Matrix (FM) (\cite{vallisneri/2008}). From the GW sign emitted by a BS, one can derive the uncertainties related to the sign parameters. In the Fourier space, one can write the GW amplitude $h$ (with $c=G=1$) as (\cite{poisson/1995}):

\begin{equation}\label{eqn:4}
\tilde{h}(f)=\mathcal{A}f^{-7/6}e^{i\psi (f)}.
\end{equation}
In the above equation, $\mathcal{A}\propto \tilde{\mathcal{M}}^{5/6}Q/D$, $\tilde{\mathcal{M}}=\eta ^{3/5}\tilde{M}$ is the redshifted chirp mass, $Q$ is a function which depends on the angles related to the position coordinates of the BS in the sky, the inclination of the orbit and polarization, $D$ is the distance to the binary,

\begin{equation}\label{eqn:5}
\psi (f)=2\pi ft_c-\phi _c-\frac{\pi}{4}+\frac{3}{128\eta v^5}\sum _{k=0}^2\alpha _kv^k,
\end{equation}
with $t_c$ representing the coalescence time, $\phi_c$ the phase of the GW when $t\rightarrow t_c$, $v\equiv(\pi\tilde{M}f)^{1/3}$ is the post-newtonian expansion parameter, which we take up to the first order $(k=2)$, $\alpha _0=1$, $\alpha _1=0$ and $\alpha _2=(20/9)(743/336+11\eta/4)$. From Eqs.(\ref{eqn:4})-(\ref{eqn:5}), the GW signal of a coalescing BS depends on:  $A$, $t_c$, $\phi_c$, $\tilde{\mathcal{M}}$ and $\eta$. However, it is convenient to calculate and to interpretate the FM in terms of the dimensionless parameters: $\overrightarrow{\Theta}\equiv lnA$, $f_0t_c$, $\phi_c$, $ln\tilde{\mathcal{M}}$ and $ln\eta$, with $f_0$ being some fiducial frequency. The FM is obtained from the derivation of $\tilde{h}(f)$ with respect to $\overrightarrow{\Theta}$. That is:

\begin{equation}\label{eqn:6}
\Gamma _{ab}=2\int _0^{f_{LSO}}\frac{\tilde{h}^*_a(f)\tilde{h}_b(f)+\tilde{h}_a(f)\tilde{h}_b^*(f)}{S_h(f)}df.
\end{equation}
In the above equation, $\tilde{h}_a(f)$, with '$a$' running through the five parameters $\overrightarrow{\Theta}$, representing partial derivatives of $\tilde{h}(f)$ with respect to $\overrightarrow{\Theta}$, and $\tilde{h}^*_a(f)$ are the derivative complex conjugates. Notice that the dependence of Eq.(\ref{eqn:6}) on $S_h(f)$ shows that the FM and posteriorly the uncertainties on the star masses depend on the GW detector sensitivity. The uncertainty of the parameters $\Theta ^i$, with $i$ running through each value of $\overrightarrow{\Theta}$, is given by $\sqrt{\langle(\Delta\Theta ^i)^2\rangle}=\sqrt{\Sigma ^{ii}}$, and ${\bf\Sigma}\equiv{\bf\Gamma}^{-1}$ is the covariance Fisher matrix (CFM), given simply by the inverse of the FM. Following \cite{cutler/1994} we can write the uncertainties on $m_1$ and $m_2$ as:

\begin{eqnarray}
\Delta m_1=\Sigma^{\mu\mu}\left[\frac{M(\mu-3m_1)}{2\mu(m_1-m_2)}\right]^{2},\\
\Delta m_2=\Sigma^{\mu\mu}\left[\frac{M(\mu-3m_2)}{2\mu(m_1-m_2)}\right]^{2},
\end{eqnarray} 
with $\mu=\eta M$. Therefore, evaluating Eqs.(6) and (7), we can ensure that the values of $m_1$ and $m_2$ lie in the ranges $1.228\,{\rm M}_\odot\leq m_1\leq 1.366\,{\rm M}_\odot$ and $0.670\,{\rm M}_\odot\leq m_2\leq 0.737\,{\rm M}_\odot$ for the aLIGO FM, while $1.272\,{\rm M}_\odot\leq m_1\leq 1.327\,{\rm M}_\odot$ and $0.687\,{\rm M}_\odot\leq m_2\leq 0.714\,{\rm M}_\odot$ for the ET FM. 

Note that, besides the range of values that the stars might achieve, the GW sign amplitude of BSs containing SSs presents some features that distinguish them from the one emitted by NSBSs, as quoted by (\cite{baus/2010}). Comparing the GW sign for such binaries, one finds that the maximal frequencies reached during the inspiral are higher for BSs with SSs. The lower values obtained for NSBSs are reasonable considering the lower compactness of the initial stars compared to the SSs. The amplitude right after the merging is higher when SSs are present, furthermore, the ring-down signal after the merging decays more rapidly for such systems. In particular, we show that from the evaluation of Eqs.(6)-(7), aLIGO and ET can constrain with high accuracy the masses involved in an NS-SS BS. However, do these systems actually exist in the Universe? We will discuss this question in the next section.

\section{What were the progenitors of GRB 070201 and GRB 051103?}

The most likely progenitor of sGRBs is thought to be a coalescence NSBS or NS-BH BS. However, \cite{abbott/2008} have excluded these systems with masses in the range ${\rm M}_\odot<m_1<3\,{\rm M}_\odot$ and ${\rm M}_\odot<m_2<40\,{\rm M}_\odot$ at $>99\%$ confidence for the GRB 070201 progenitor, while \cite{abadie/2012} have excluded an NS-BH merger with $>99\%$ confidence for GRB 051103. Both papers show a disability in constraining the GRB progenitors since the bursts of GWs were not detected. As possible explanations to these non-detections, the authors argue that since up to $15\%$ of sGRBs might be giant flares from soft gamma repeaters (SGRs) (NSs with extremely large magnetic fields), then those sources might actually be such flares. In fact, this possibility is also presented in \cite{frederiks/2007}, \cite{mazets/2008}, \cite{ofek/2008}, and \cite{hurley/2010}. Furthermore, the sensitivity of a GW detector in the case of compact binaries depends on the inclination of the orbital plane to the line of sight, the location of the source in the sky and the orbital plane relative to the line of sight.

However, if we consider that both events were due to extragalactic SGRs, the peak luminosity of these sources would be $> 10^{47}\,{\rm erg}\,{\rm s}^ {-1}$ (\cite{chapman/2009}). On the other hand, if we consider the more common short duration bursts from SGRs, the luminosities are typically $\lesssim 10^{41}\,{\rm erg}\,{\rm s}^ {-1}$. Moreover, \cite{chapman/2009} estimate the rate of flares with peak luminosity $> 10^{47}\,{\rm erg}\,{\rm s}^ {-1}$ to be to rather rare. Adopting their flare rate, the probability of observing two such flares within $5\,{\rm Mpc}$ during the $17\,{\rm yr}$ of IPN3\footnote{The third interplanetary network (IPN3) is a group of spacecraft equipped with GRB detectors.} observation is $\sim 1\%$. An alternative to explain the absence of GW signal from these GRBs arises if we consider that at least one of the stars of the BS is an SS with mass $M\lesssim 0.7{\rm M}_\odot$. Beyond that, during the merger process, but before the final plunge, SM of the SS could be accreted by the NS. Once there is a seed of SM inside a NS, this strange seed will absorb neutrons, protons, and hyperons (if they are present in the NS), liberating their constituent quarks. The conversion of the whole NS will then occur in a very short time, typically $1\, {\rm ms}-1\,{\rm s}$, across a detonation mode similar to that observed in GRBs (\cite{horvath/1988}).

In terms of electromagnetic emission for GRB 070201, the values of the fluence and peak flux correspond to an isotropic energy output $\sim 10^{45}\,{\rm erg}$ and an isotropic peak luminosity $\sim 10^ {47}\,{\rm erg}\,{\rm s^ {-1}}$ if the source of GRB 070201 is situated in M31 at $D=780$kpc. Returning to the concerned NS-SS BS presented in Section 2, it could reach frequencies significantly higher than an NSBS. In particular, the NS-SS BS would not be detected by initial LIGO, by using matched filtering technique, only if the inclination angle\footnote{The angle between the orbital plane and the vector connecting the origin of the detector to the origin of the source.} $\iota$ is such that $\cos\iota<0.80$ for GRB 051103 and $\cos\iota<0.20$ for GRB 070201 (considering that the NS-SS components are, respectively, $m_{1}=1.3\,{\rm M}_\odot$ and $m_{2} = 0.7\,{\rm M}_\odot$). If one considers the SS with $0.07{\rm M}_\odot$, then GRB 070201 and GRB 051103 would not be detected even when $\cos\iota=1$. Thus, there would be no detection. Recently, \cite{arun/2014} have discussed a method that could measure the inclination angle to an accuracy of $\sim 5.1$ degrees for an NSBS system at 200 Mpc if the direction of the source and the redshift are known electromagnetically. This method can be configured into a powerful way of studying compact binaries (e.g., NS-NS, NS-BH, NS-SS, SS-SS) in the era of advanced GW detectors.

It is important to mention that the matched filtering combines the definition of the signal-to-noise ratio with a template normalization function $\sigma_{\rm h}$. Then, an effective distance is defined for a given trigger as $D_{\rm eff}=\sigma_{\rm h}/\rho_{0}$. In this case, $D_{\rm eff}$ is related to the distance from the source ($D$), the inclination angle ($\iota$), and the detector antenna patterns\footnote{The antenna patterns of the detector give the sensitivity to the polarizations of GW signals.}. In particular, some of the parameters used by LIGO team in the analysis of compact binary sources are: (i) the total mass is distributed uniformly between 2 and $\sim 35-40\,{\rm M}_\odot$; (ii) the mass ratio $q = m_{1} /m_{2}$ is distributed uniformly between 1 and the maximum value such that $m_{1},m_{2} > 1\,{\rm M}_\odot$. If we take a BS composed by two stars with $1\,{\rm M}_\odot$, then the total number of cycles in the band of initial LIGO would be $\sim 2,900$ . On the other hand, for the limit case of $m_{1}-m_{2}=1.3-0.7\,{\rm M}_\odot$ we have $\sim 3,200$ cycles inside the band of the LIGO detectors. This difference in the total number of GW cycles could reduce the efficiency of the template when using the matched filtering technique\footnote{In the total accumulated phase of the wave detected in the sensitive bandwidth of the detector, the template must match the signal to a fraction of a cycle. This implies a high phasing accuracy (see, for example, \cite{will/1998}).}. Thus, a possible consistency test could come from a re-analysis of the LIGO signals for GRB 070201 and GRB 051103 taking into account some possible cases involving SSs. That is, for example, the LIGO team could use the total mass distributed between $1.3$ and $\sim 2\,{\rm M}_\odot$ that corresponds to the mass of the SS in the range $\sim 0.07-0.7\,{\rm M}_\odot$.

\section{Final Remarks}

In the present paper we have obtained the $m_1$ and $m_2$ uncertainties from the calculated FMs with the purpose of showing that in the era of advanced GW detectors it will be possible to infer, from the combination of the maximum GW frequency and the accuracy in the determination of the masses, if SSs do exist. We also discussed the possibility that the progenitors of GRB 051103 and GRB 070201 could be BSs composed of an NS and an SS. Considering a direct detection of GWs it would be unlikely that the maximum frequency of a binary consisting of two NSs ($m_1=1.3\,{\rm M}_\odot$ and $m_2=0.7\,{\rm M}_\odot$) can exceed the limit $\sim 500\,{\rm Hz}$. This is due to the high tidal forces acting on the BS. In the case of a BS consisting of an NS with mass $\sim 1.3\,{\rm M}_\odot$ and an SS with mass $\sim 0.7\,{\rm M}_\odot$ (whose radius is $\sim 7.8\,{\rm km}$ if we consider the bag constant $B=100\,{\rm MeV}\,{\rm fm}^{-3}$) it is likely that this NS-SS system achieves maximum frequency greater than that achieved by an NSBS. This happens because the GWs emitted during these events are primarily governed by the compactness ($M/R$) of each star. Thus, an NS-SS BS (with masses $m_1\sim 1.3\,{\rm M}_\odot$ and $m_2\lesssim 0.7\,{\rm M}_\odot$) could easily overcome the barrier of $\sim 1.5\,{\rm kHz}$. Furthermore, the gravitational energy released by the merger of an NS with an SS could reach $E\sim 10^{53}\,{\rm erg}$ if we consider $B=100\,{\rm MeV}\,{\rm fm}^{-3}$. Moreover, if before the final plunge the NS accretes SM from the companion SS, the conversion NS$\rightarrow$SS could liberate $E^{conv} \sim 10^{53}\,{\rm erg}$ (\cite{bombaci/2000}) in agreement with the energy required to power GRB sources at cosmological distances. Finally, an SS may have a thin baryonic crust, as discussed e.g. by \citealp{huang97}, with a typical mass $\sim 10^{-5}\,{\rm M}_\odot$. In this case, due to the fact that the amount of baryons contaminating the fireball cannot exceed the mass of this thin crust, we could have, for these systems, a fireball with high relativistic factors when compared to an NSBS with the same total mass.

Concerning the maximum value of an NS mass, we refer the reader to the results presented by \cite{kiziltan/2010} - $1.5-3.2$M$_\odot$ - which are obtained from numerically integrating the Oppenheimer-Volkoff equations for a low-density EoS at the lowest energy state of the nuclei (\cite{baym/1971}) and an EoS which includes some nuclear processes which alter their stiffness (\cite{kalogera/1996}). As was mentioned above, there is no minimum mass limit for an SS, while its maximum value depends on the specification of $B$ and other MIT parameters. In \cite{weber/2012}, this upper limit is $\sim2$M$_\odot$, while in \cite{benhar/2007}, a more embracing approach, a variation in the MIT parameters put the limit $\sim1.8$M$_\odot$. It might also be usefull here to emphasyze that the concerned less massive star is unlikely to be a `normal' WD. Since the radius of a WD is much bigger than its Schwarzschild radius, the last stable orbit of a BS containing a WD is reached when the stars touch each other. Once the radius of an NS is negligible when compared to the radius of a WD ($\sim 10^3$km), the last stable orbit frequency of an NS-WD BS depends on the inverse of the WD radius only. So, it is unlikely to have $f_{LSO}>1$Hz in BSs containing a WD (see, e.g., \cite{rosado/2011}). Therefore when working with a BS composed of at least one WD, it is expected to have a last stable orbit frequency widely different from what was obtained in our case. From \cite{rodrigues/2011}, one can see that for different values of the EoS parameters, $R_{SS}<10$km for $m_{SS}=0.7$M$_\odot$, so the tidal effects on such a star would be considerably lower than on a WD. In this way, Eq.(1) successfully defines the last stable orbit of the concerned BS.

One might ponder the reason which allows us to assume the SS mass $m_2=0.7\,{\rm M}_\odot$ as our upper limit. In this framework, let us remind the reader that the search for massive compact halo objects (MACHOs) has showed that $\sim20\%$ of the Galactic halo is populated by compact objects with masses in the range $0.15-0.9\,{\rm M}_\odot$ (\cite{alcock/2000}). A similar result was obtained by \cite{cauchi/2005} who concluded that the average mass of MACHOs lies in the range $0.5-1\,{\rm M}_\odot$. Note that a non-negligible portion of these objects could not be WDs since it would imply a high luminosity for the halo. In this way, at least a certain fraction of these halo objects could be, for instance, SSs. Let us consider a standard power-law mass function $\phi(m)\propto m^{-\alpha}$. For the $\alpha$ values we refer the reader to the works of, e.g., \cite{han/1996,besla/2013}, among others, who have determined $\alpha \sim 1.3-2.35$ for MACHOs in the range $0.075-1\,{\rm M}_\odot$. Given this range for the values ​​of $\alpha$, the fraction of compact low-mass objects in the halo of Milky Way, with a mass close to $0.7\,{\rm M}_\odot$, will be $f \sim 0.25-0.60$ per cent. A substantial part of these MACHOs might be SSs with $m\sim 0.7\,{\rm M}_\odot$. Although we have used the halo of the Milky Way for that estimate, it is likely that compact objects with a mass $\lesssim 0.7\,{\rm M}_\odot$, as SSs, also exist in the bulge and disk, as well as in other galaxies.

Concerning the advanced detectors and, in particular, from the evaluation of Eq.(\ref{eqn:3}), one would expect $\sim2$ NS-SS coalescence events to be detected by ET in each day. Once the ET is operating, if this event day rate do not be satisfied, one can infer the NS-SS Galactic merger rate to be $\mathcal{R}^{NS-SS}< 10^{-5}$yr$^{-1}$. Concerning aLIGO, one NS-SS coalescence event is expected in each $\sim 8\,{\rm yr}$. Whether SM does exist in nature is still an open question. Although it is unlikely that the answer to this question will come from laboratory experiments, due to the fact that quarks are not observable as individual particles, among other reasons. In this way, GW detectors might bring up some evidence for the existence of SSs and consequently confirm the absolute ground state of matter to be the state of SM.

\section{Acknowledgments}
PHRSM would like to thank CAPES for financial support. ODM would like to thank CNPq for partial financial support (grant 304654/2012-4). We would like to thank the referee for his/her detailed comments that have improved the paper.

\label{lastpage}


\begin{thebibliography}{}

\bibitem[\protect\citeauthoryear{Abadie et al.}{2010}]{abadie/2010} Abadie J. et al., 2010, Classical Quantum Gravity, 27, 173001.
\bibitem[\protect\citeauthoryear{Abadie et al.}{2012}]{abadie/2012} Abadie J., 2012, Astrophys. J., 755, 2.
\bibitem[\protect\citeauthoryear{Abbott et al.}{2008}]{abbott/2008} Abott B., 2008, Astrophys. J., 681, 1419.
\bibitem[\protect\citeauthoryear{Akmal 1998}{}]{akmal/1998} Akmal A., 1998, Phys. Rev. C, 58, 1804. 
\bibitem[\protect\citeauthoryear{Alcock et al. 2000}{}]{alcock/2000} Alcock C. et al., 2000, Astrophys. J., 542, 281.
\bibitem[\protect\citeauthoryear{Arun et al.}{2005}]{arun/2005} Arun K.G. et al., 2005, Phys. Rev. D, 71, 084008.
\bibitem[\protect\citeauthoryear{Arun et al.}{2014}]{arun/2014} Arun K.G., Tagoshi H., Pai A., Mishra C.K., 2014, {\bf arXiv}:1403.6917
\bibitem[\protect\citeauthoryear{Bauswein et al.}{2009}]{bauswein/2009} Bauswein A. et al., 2009, Phys. Rev. Lett., 103, 011101.
\bibitem[\protect\citeauthoryear{Bauswein et al. 2010}{}]{baus/2010} Bauswein A., Oechslin R. and Janka H.-T., 2010, Phys. Rev. D., 81, 024012 
\bibitem[\protect\citeauthoryear{Baym et al. 1971}{}]{baym/1971} Baym G. et al., 1971, Astrophys. J., 170, 299.
\bibitem[\protect\citeauthoryear{Bejger et al. 2005}{}]{bejger/2005} Bejger M., 2005, Astron. Astrophys., 431, 297.
\bibitem[\protect\citeauthoryear{Belczynski et al. }{2002}]{belczynski/2002} Belczynski K., Bulik T, Klu{\'z}niak W., 2002, Astrophys. J., 567, L63.
\bibitem[\protect\citeauthoryear{Belczynski et al. }{2008}]{belczynski/2008} Belczynski K. et al., 2008, Astrophys. J., 680, L129.
\bibitem[\protect\citeauthoryear{Benhar et al.}{2007}]{benhar/2007} Benhar O. et al., 2007, Gen. Rel. Grav., 39, 1323.
\bibitem[\protect\citeauthoryear{Berger }{2010}]{berger/2010} Berger E., 2010, Astrophys. J., 722, 1946.
\bibitem[\protect\citeauthoryear{Besla }{2013}]{besla/2013} Besla G., Hernquist L., Loeb A., 2013, MNRAS, 428, 2342.
\bibitem[\protect\citeauthoryear{Blinnikov 1984}{}]{blinnikov/1984} Blinnikov S.I., 1984, Soviet Astronomy Letters, 10, 177.
\bibitem[\protect\citeauthoryear{Bodmer 1971}{}]{bodmer/1971} Bodmer A.R., 1971, Phys. Rev. D, 4, 1601.
\bibitem[\protect\citeauthoryear{Bombaci and Datta 2000}{}]{bombaci/2000} Bombaci I. and Datta B., 2000, Astrophys. J., 530, L69
\bibitem[\protect\citeauthoryear{Cauchi Novati et al.}{2005}]{cauchi/2005} Calchi Novati S. et al., 2005, A\&A, 443, 911
\bibitem[\protect\citeauthoryear{Chapman et al. 2009}{}]{chapman/2009} Chapman R., Priddey R.S., Tanvir N.R., 2009, MNRAS, 395, 1515 
\bibitem[\protect\citeauthoryear{Coburn and Boggs 2003}{}]{coburn/2003} Coburn W.E. and Boggs S.E., 2003, Nature, 423, 415.
\bibitem[\protect\citeauthoryear{Corvino et al.}{2012}]{corvino/2012} Corvino G. et al., 2012, arXiv:astro-ph/1203.5110. 
\bibitem[\protect\citeauthoryear{Coward }{2007}]{coward/2007} Coward D., 2007, New Astronomy Reviews, 51, 539.
\bibitem[\protect\citeauthoryear{Cutler et al. 1993}{}]{cutler/1993} Cutler C. et al., 1993, Phys. Rev. Lett., 70, 2984.  
\bibitem[\protect\citeauthoryear{Cutler and Flanagan}{1994}]{cutler/1994} Cutler C. and Flanagan {\'E}.E., 1994, Phys. Rev. D, 49, 2658.  
\bibitem[\protect\citeauthoryear{Faber et al. 1993}{}]{faber/2002} Faber J.A. et al., 2002, Phys. Rev. Lett., 89, 231102.
\bibitem[\protect\citeauthoryear{Fiore et al. 2006}{}]{fiore/2006} Fiore M., 2006, Mon. Not. R. Astron. Soc., 372, 1851.
\bibitem[\protect\citeauthoryear{Frederiks et al.}{2007}]{frederiks/2007} Frederiks D.D., 2007, Astronomy Letters, 33, 19.
\bibitem[\protect\citeauthoryear{Glendenning 1995}{}]{glendenning/1995a} Glendenning N.K., 1995, Phys. Rev. Lett., 74, 3519.
\bibitem[\protect\citeauthoryear{Glendenning et al. 1995}{}]{glendenning/1995b} Glendenning N.K. et al., 1995, Astrophys. J., 450, 253.
\bibitem[\protect\citeauthoryear{Guillemot 2012}{}]{guillemot/2012} Guillemot L., 2012, Mon. Not. R. Astron. Soc., 422, 1294.
\bibitem[\protect\citeauthoryear{Haensel et al.}{1986}]{haensel/1986} Haensel P., 1986, Astron. Astrophys., 160, 121.
\bibitem[\protect\citeauthoryear{Hamada and Salpeter 1961}{}]{hamada/1961} Hamanda, T. and Salpeter, E.E., 1961, Astrophys. J., 134, 683.
\bibitem[\protect\citeauthoryear{Han and Gould}{1996}]{han/1996} Han C., Gould A., 1996, Astrophys. J., 467, 540. 
\bibitem[\protect\citeauthoryear{Harry et al. 2010}{}]{harry/2010} Harry G.M. et al., 2010, Classical Quantum Gravity, 27, 084006.
\bibitem[\protect\citeauthoryear{Horvath and Benvenuto 1988}{}]{horvath/1988} Horvath, J.E. and Benvenuto, O.G., 1988, Phys. Lett. B,  213,516
\bibitem[\protect\citeauthoryear{Huang and Lu 1997}{}]{huang97} Huang Y.F. and Lu T., 1997, A\&A, 325, 189
\bibitem[\protect\citeauthoryear{Hurley et al.}{2010}]{hurley/2010} Hurley K. et al., 2010, Mon. Not. R. Astron. Soc., 403, 342.
\bibitem[\protect\citeauthoryear{Kalogera et al.}{2004}]{kalogera/2004} Kalogera V., 2004, Astrophys. J., 601, L179. Erratum-ibid., 2004, 614, L137. 
\bibitem[\protect\citeauthoryear{Kalogera and Baym 1996}{}]{kalogera/1996} Kalogera V. and Baym G., 1996, Astrophys. J., 470, L61.
\bibitem[\protect\citeauthoryear{Kiziltan}{2010}]{kiziltan/2010} Kiziltan B., 2010, arXiv:astro-ph/1011.4291. 
\bibitem[\protect\citeauthoryear{Kopparapu et al. 2008}{}]{kopparapu/2008} Kopparapu R.K. et al., 2008, Astrophys. J., 675, 1459.
\bibitem[\protect\citeauthoryear{Koshy et al. 2012}{}]{koshy/2012} Koshy S. et al., 2011, 2011 Annual Meeting of the California-Nevada Section of the APS, 56, B2006.
\bibitem[\protect\citeauthoryear{Kouveliotou 1993}{}]{kouveliotou/1993} Kouveliotou C., 1993, Astrophys. J., 413, L101.
\bibitem[\protect\citeauthoryear{Lattimer 2012}{}]{lattimer/2012} Lattimer J.M., 2012, Ann. Rev. Nuc. Part. Sci., 62, 485.
\bibitem[\protect\citeauthoryear{Mathews et al.}{2006}]{mathews/2006} Mathews G.J. et al., 2006, Journal of Physics G Nuclear Physics, 32, 747.
\bibitem[\protect\citeauthoryear{Mazets et al.}{2008}]{mazets/2008} Mazets E.P. et al., 2008, Astrophys. J., 680, 545.
\bibitem[\protect\citeauthoryear{Medvedev and Loeb 1999}{}]{medvedev/1999} Medvedev M.V. and Loeb A., 1999, Astrophys. J., 526, 697.
\bibitem[\protect\citeauthoryear{Medvedev et al. 2005}{}]{medvedev/2005} Medvedev M.V. et al., 2005, Astrophys. J., 618, L75. 
\bibitem[\protect\citeauthoryear{Mishra et al.}{2010}]{mishra/2010} Mishra C.K., 2010, Phys. Rev. D, 82, 064010.
\bibitem[\protect\citeauthoryear{Nakar 2007}{}]{nakar/2007} Nakar E., 2007, Phys. Rep., 442, 166.
\bibitem[\protect\citeauthoryear{Oechslin et al. 2012}{}]{oechslin/2004} Oechslin K., 2004, Mon. Not. R. Astron. Soc., 349, 1469.
\bibitem[\protect\citeauthoryear{Ofek et al.}{2008}]{ofek/2008} Ofek E.O. et al., 2008, Astrophys. J., 681, 1464.
\bibitem[\protect\citeauthoryear{Phinney 1991}{}]{phinney/1991} Phinney E.S., 1991, Astrophys. J., 380, L17.
\bibitem[\protect\citeauthoryear{Poisson and Will 1995}{}]{poisson/1995} Poisson E. and Will C.M., 1995, Phys. Rev. D, 52, 848.
\bibitem[\protect\citeauthoryear{Provencal et al.}{1998}]{provencal/1998} Provencal J.L. et al., 1998, Astrophys. J., 494, 759.
\bibitem[\protect\citeauthoryear{Provencal et al.}{2002}]{provencal/2002} Provencal J.L. et al., 2002, Astrophys. J., 568, 324.
\bibitem[\protect\citeauthoryear{Regimbau and de Freitas Pacheco}{2006}]{regimbau/2006} Regimbau T. and de Freitas Pacheco J.S., 2006, Astrophys. J., 642, 455. 
\bibitem[\protect\citeauthoryear{Rodrigues et al.}{2011}]{rodrigues/2011} Rodrigues H. et al., 2011, Astrophys. J., 730, 31.
\bibitem[\protect\citeauthoryear{Rosado}{2011}]{rosado/2011} Rosado P.A., 2011, Phys. Rev. D, 84, 084004. 
\bibitem[\protect\citeauthoryear{Sathyaprakash 2001}{}]{sathyaprakash/2001} Sathyaprakash B.S., 2001, Pramana, 56, 457.
\bibitem[\protect\citeauthoryear{Sathyaprakash et al. 2012}{}]{sathyaprakash/2012} Sathyaprakash B. et al., 2012, Classical Quantum Gravity, 29, 124013.
\bibitem[\protect\citeauthoryear{Taniguchi and Gourgoulhon 2003}{}]{taniguchi/2003} Taniguchi K. and Gourgoulhon E., 2003, Phys. Rev. D, 68, 124025.
\bibitem[\protect\citeauthoryear{Vallisneri 2008}{}]{vallisneri/2008} Vallisneri M., 2008, Phys. Rev. D, 77, 042001. 
\bibitem[\protect\citeauthoryear{Waxman 2003}{}]{waxman/2003} Waxman E., 2003, Nature, 423, 388.
\bibitem[\protect\citeauthoryear{Weber 2006}{}]{weber/2006} Weber F., 2006, Int. J. Mod. Phys. D, 11, 2.
\bibitem[\protect\citeauthoryear{Weber 2007}{}]{weber/2007} Weber F., 2007, arXiv:astro-ph/0705.2708.
\bibitem[\protect\citeauthoryear{Weber et al.}{2012}]{weber/2012} Weber F. et al., 2012, Proceedings IAU Symposium, 291.
\bibitem[\protect\citeauthoryear{Weibel 1959}{}]{weibel/1959} Weibel E.S., 1959, Phys. Rev. Lett., 2, 83. 
\bibitem[\protect\citeauthoryear{Will}{1998}]{will/1998} Will C.M., 1998, ECONFC9808031:02, Lecture notes from the 1998 Slac Summer Institute on Particle Physics (arXiv:gr-qc/9811036).
\bibitem[\protect\citeauthoryear{Witten 1984}{}]{witten/1984} Witten E., 1984, Phys. Rev. D, 30, 272.
\bibitem[\protect\citeauthoryear{Xu 2005}{}]{xu/2005} Xu R., 2005, Ap. \& S.S., 297, 179.
\bibitem[\protect\citeauthoryear{Zhang and M\'es\'zaros 2004}{}]{zhang/2004} Zhang B. and M\'es\'zaros P., 2004, Int. J. Mod. Phys. A, 19, 2385.

\end{thebibliography}
\end{document}